\def\be{\begin{equation}}
\def\ee{\end{equation}}

\documentclass[%
aps,%
prb,%
twocolumn,%preprint,%
superscriptaddress,%
showpacs,%
preprintnumbers,%
byrevtex%
]{revtex4}
\usepackage{epsfig}

\newcommand{\labs} {\left\vert}
\newcommand{\rabs} {\right\vert}

\newcommand{\lab} {\left\langle}
\newcommand{\rab} {\right\rangle}

\usepackage{color}
\begin{document}
\title[Theoretical studies of spin-dependent electronic transport in ferromagnetically contacted graphene flakes]
      {Theoretical studies of spin-dependent electronic transport in ferromagnetically contacted graphene flakes}
\author{\bf S. Krompiewski}
\address{Institute of Molecular Physics, Polish Academy of
Sciences, ul.~M.~Smoluchowskiego 17, 60179 Pozna{\'n}, Poland}
%\author{\bf A.K.B}
%\address{}

\date{\today}

\begin{abstract}

Based on a tight-binding model and a recursive Green's function
technique, spin-depentent ballistic transport through tinny
graphene sheets (flakes) is studied. The main interest is focussed
on: electrical conductivity, giant magnetoresistance (GMR) and
shot noise. It is shown that when graphene flakes are sandwiched
between two ferromagnetic electrodes, the resulting GMR
coefficient may be quite significant. This statement holds true
both for zigzag and armchair chiralities, as well as for different
aspect (width/length) ratios. Remarkably, in absolute values the
GMR of the armchair-edge graphene flakes is systematically greater
than that corresponding to the zigzag-edge graphene flakes. This
finding is attributed to the different degree of conduction
channel mixing for the two chiralities in question. It is also
shown that for big aspect ratio flakes, 3-dimensional
end-contacted leads, very much like invasive contacts, result in
non-universal behavior of both conductivity and Fano factor.
\end{abstract}
%\jl{3}
%\ead{stefan@ifmpan.poznan.pl}
\pacs{81.05.Uw, 75.47.De, 75.47.Jn}
% 81.05.Uw, 75.47.De, 73.23.Ad
% Carbon, diamond, graphite; GMR, Ballistic magnetoresistance
\maketitle

\section{Introduction}

Recently a lot of interest has been directed to carbon-based
systems, in search for alternative materials which would make it
possible to go beyond the silicon technology. While looking at
history of studies along this line, one can point out two
important milestones: \emph{(i)} discovery of carbon nanotubes
(CNTs), honeycomb-lattice cylinders, dating back to 1991,
\cite{Ijima} and \emph{(ii)} fabrication of individual atomic
planes, called graphene (Gr), by exfoliation from graphite in
2004.\cite{Novoselov} So far, on obvious grounds, the CNTs have
been much more thoroughly investigated than Gr, but this
difference diminishes very quickly. This paper focusses on spin
transport problems, mostly on giant magnetoresistance, related to
potential applications of graphitic nanostructures in spintronics.
In this respect, quite a lot has been done in the case of CNTs,
there are hundreds of both
experimental\cite{Tsuka,Schoe,Hauptmann,Liang,Sapmaz} and
theoretical\cite{mehrez,ustron,krompiewski06,krompiewskiNANO}
papers, covering all the transport regimes (ballistic, Coulomb
blockade, Schottky barrier and Kondo) which show that the GMR or
TMR (T for tunnel) effects are usually quite considerable. The
respective studies on Gr are still scarce. The pioneering
experimental paper on magnetoresistance is Ref[\onlinecite{Hill}],
where Gr spin valve devices with permalloy contacts have been
shown to have the GMR effect $\simeq$ 10\% at room temperature.
The following experimental paper,\cite{Cho} reports on conductance
of Gr showing Fabry-Perot-like patterns, and pronounced
oscillations of GMR (including changes in sign). From the
theoretical point of view, the debate is still on, and the
question whether or not the Gr-based spin valves have a good
performance is still open. The results published so far show that
the answer to this question depends critically on the contacts
(cf. Ref.[\onlinecite{Brey}] and
Refs.[\onlinecite{Kim-Kim,Ding}]). In this study, in contrast to
those reported hitherto by other theoreticians, a 3-dimensional
contacts are used.

The paper is organized as follows: In Sec.~II the theoretical
method based on a tight-binding model, as well as the way the
graphene sheets and the contacts have been modelled, are shortly
outlined. Sec.~III is devoted to presentation of results, whereas
the subsequent section summarizes the main results.

%%%%%%%%%%%%%%%%%%%%%%%%%%%%%%%%%%%%%%%%%%%%%%%%%%%%%%%%%%%%%%%%%%%%%%%%%%%%%
\begin{figure}[t]
\epsfxsize=6truein                      % <--- sets horizontal size
\centerline{\epsfbox{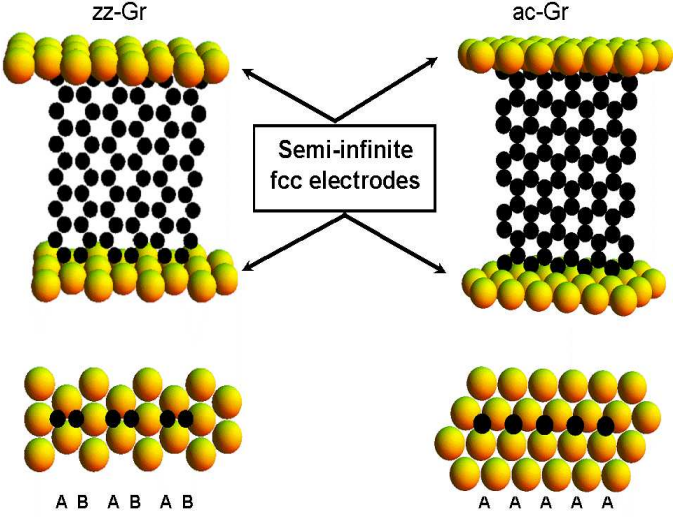}}   % <--- centers your figure
%\centering \epsfxsize=3.2in \epsfysize=2.5in
%\rotatebox{0}{\epsfbox{obrazy_ac-zz.eps}}
\caption{ (Color online) Graphene sheets attached to semi-infinite
3-dimensional metallic electrodes.  The left and right parts
correspond to the zigzag- and armchair-edge ribbons, respectively.
The lower part shows details of the interfaces, whereas the
letters A, B denote a type of the sublattice the interface carbon
atoms belong to. } \label{scheme}
\end{figure}
%%%%%%%%%%%%%%%%%%%%%%%%%%%%%%%%%%%%%%%%%%%%%%%%%%%%%%%%%%%%%%%%%%%%%%%%%%%%%%

\section{Method and Modelling}

The method employed here is similar to that used earlier while
dealing with carbon nanotubes inserted between ferromagnetic
contacts,\cite{ustron,krompiewski06,krompiewskiNANO} except that
carbon allotrope of interest now is graphene . The single
$\pi$-orbital tight-binding Hamiltonian describes a graphene sheet
of width W and length L (in the current direction). The
semi-infinite metallic electrodes extend from $<-\infty,0>$ and
$<L,\infty>$ for the source and drain, respectively.
 The ferromagnetic electrodes are supposed to have spin-split
 s-bands, mimicking d-bands of real transition metals.
 The total Hamiltonian reads

\be \label{H} H = - \sum \limits_{ i ,j, \sigma } t_{i,j}\labs i,
\sigma \rab \lab \sigma, j \rabs +\sum \limits_{i, \sigma}
\epsilon_{i, \sigma} \labs i, \sigma \rab \lab \sigma, i \rabs,
%\end{equation}
\ee where i and j run over the whole device (i.e graphene and the
electrodes), $\sigma$ is the spin index, and $t_{i,j}$ and
$\epsilon_{i,\sigma}$ stand for the hopping integrals and the
on-site potentials, respectively. The systems under study are
impurity-free with well transparent interfaces (strong coupling
limit) so neglecting of correlations in the Hamiltonian should be
justified if one restricts oneself just to the ballistic transport
regime, putting aside the Coulomb blockade\cite{Liang,Weymann} and
Kondo \cite{Sapmaz,Hauptmann} regimes. In this regard, the on-site
parameters in the present approach are used to take into account
the effect of the gate voltage in the graphene sheet as well as
the spin band-splitting in the metallic electrodes.

Fig. \ref{scheme} shows schematically devices of the present
interest, \emph{viz.} zigzag-edge graphene (zz-Gr) and
armchair-edge graphene (ac-Gr). The graphene sheets (black
spheres) are sandwiched between metallic contacts (light spheres).
The diameters of the spheres correspond to the nearest neighbor
spacings. The unit cells in the vertical direction are the blunt
saw-teeth lines for zz-Gr, and double zigzag lines for ac-Gr. As
readily seen, it is assumed that there is a perfect lattice
matching between graphene and the electrodes. In fact this
assumption is acceptable since the graphene lattice constant
($a_{Gr}=2.46 \AA$) fits really well to interatomic distances in
such metals like: Cu (2.51 \AA), Ni (2.49 \AA) or Co (2.55
\AA).\cite{Karpan}

The use of a single-orbital electrodes allows to write an analytic
expression for the electrodes surface Green functions in ${\bf
k}$-space.
 \begin{eqnarray} \label{SGF}
g_\sigma ({\bf k},E)&=&\frac{E -\epsilon_\sigma({\bf k}) \pm
\sqrt{(E -\epsilon_\sigma({\bf k}))^2-4|w({\bf k})|^2 } }
{2|w({\bf k})|^2 } \nonumber \\
\epsilon_\sigma({\bf k})&=&2t (\cos{k_x a}+2\cos{\frac{k_x a}{2}} \cos{ \frac{\sqrt{3}k_y a}{2} } )+\Delta_\sigma \nonumber \\
 w({\bf k})&=& -t (2\cos({\frac{k_x a}{2}}) e^{{\frac{i k_{y} a} { 2\sqrt{3} }} }
+e^{{-\frac{ik_y a}{\sqrt{3}}}}), % \nonumber
\end{eqnarray}
where $a=\sqrt 2 a_{Gr}$ is the fcc-metal lattice constant,
$g_\sigma$ is the surface Green function for spin $\sigma$, and
$\Delta_\sigma$ is a rigid band splitting chosen so as to give a
desirable spin polarization P of the electrodes. Here P=50\% has
been set, corresponding to $\Delta_\uparrow=-2.32 \, t$ and
$\Delta_\downarrow=1.6 \, t$ (in the paramagnetic case
$\Delta_\sigma=-0.86 \, t$). Incidentally, this  simple
parametrization has already been shown to work satisfactorily well
in the case of carbon nanotube/ferromagnet
systems.\cite{krompiewski06,ustron} $\epsilon_\sigma({\bf k})$ in
Eq. (2) is the fcc(111)-surface energy spectrum, calculated for a
semi-infinite metal slab according to the method described in
Ref.~[\onlinecite{Todorov}]. The surface k-vectors lie on the
metal surface but upon attachment of graphene they are no longer
good quantum numbers, so the trick is to perform a Fourier
transformation to the real space and work with those surface Green
function which are close to the graphene interface. After having
transformed the surface Green function to the real-space, the
self-energies $\Sigma_\sigma ^\alpha$ and the corresponding
spectral functions $\Gamma_\alpha ^\sigma$ are computed from
$\Sigma_\alpha ^\sigma=T g_\sigma ^\alpha T^\dagger$ and $
\Gamma_\alpha ^\sigma = i ( \Sigma_\alpha ^\sigma -{\Sigma_\alpha
^\sigma} ^\dagger ) $, respectively. With $\alpha=$L or R,
referring the left and right electrodes, and T being the
Gr/electrode coupling matrices. Henceforth, the spin indexes
$\sigma$ will be skipped for brevity.

The recursive method goes as follows \cite{JPCM04}
\begin{eqnarray} \label{aaa}
g_L(0) &\equiv& g_L, \; \;  g_R(N+1) \equiv g_R , \nonumber \\
g_{L,R}(i)&=&(E-D_i-\Sigma_{L,R}(i))^{-1} \nonumber \\
\Sigma_L(i)&=&T_{i,i-1} g_L(i-1) T_{i-1,i}, \; \; \nonumber \\
\Sigma_R(i)&=&T_{i,i+1} g_R(i+1) T_{i+1,i},
\end{eqnarray}
\begin{equation} \label{G}
G_{i}=(E-D_i-\Sigma_L(i)-\Sigma_R(i))^{-1}.
\end{equation}
Above, $g_{L,R}$ are local Green's functions for the i-th unit
cell of graphene, the matrices D and T stand for the diagonal and
off--diagonal Hamiltonian sub-matrices, whereas the full Green's
function is given by Eq.~\ref{G}. The graphene unit cells extends
from i=1 (following the L electrode) up to i=N (preceding the R
electrode). So the recursion starts with the metal-interface Green
functions $g_L(0)$ and $g_R(N+1)$ being Fourier components of the
surface Green functions defined in Eq.(\ref{SGF}).

The other quantities of interest are transmission ($\cal T$),
conductivity ({\Large $\sigma$}) , and shot noise Fano factor (F),
as well as giant magnetoresistance (GMR). In the ballistic
transport regime and at zero temperature these quantities read:

%\begin{equation} \label{T} {\cal T}=Tr(\Gamma_{i}^L G_{i}
%\Gamma_{i}^R G_{i}^{\dagger})
%\end{equation}

\begin{eqnarray} \label{GMR}
\mathcal{T} &=& \Gamma_{i}^L G_{i} \Gamma_{i}^R G_{i}^{\dagger},
\nonumber \\
 \textrm{\Large $\sigma$}  &=&(L/W) \frac{e^2}{h} Tr[{\cal T}(E_F)], \nonumber
 \\ [2mm]
 F & = &  Tr[\mathcal{T}(E) \left(1-\mathcal{T}(E) \right)]/Tr [\mathcal{T}(E)] , \nonumber\\
GMR&=&100 (1- \textrm{\Large
$\sigma$}_{\uparrow,\downarrow}/\textrm{\Large
$\sigma$}_{\uparrow,\uparrow}),
\end{eqnarray}

where the arrows $\uparrow \uparrow$ and $\uparrow \downarrow$
denote parallel and antiparallel alignments of ferromagnetic
electrodes.
%%%%%%%%%%%%%%%%%%%%%%%%%%%%%%%%%%%%%%%%%%%%%%%%%%%%%%%%%%%%%%%%%%%%%%%%%%%%%%
\begin{figure}[t]
\centering \epsfxsize=3.0in \epsfysize=3.2in
\rotatebox{0}{\epsfbox{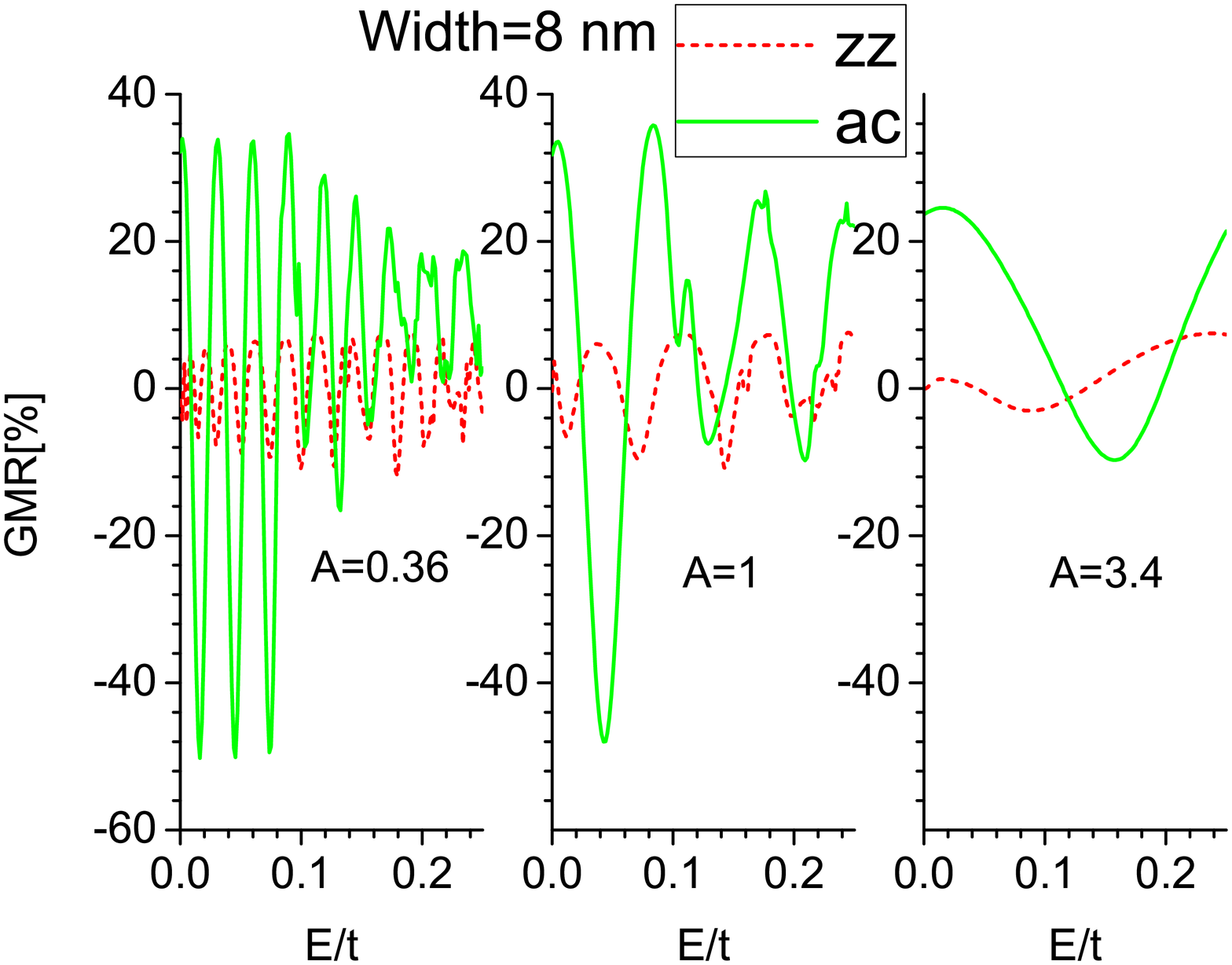}} \caption{ (Color online) Giant
magnetoresistance for ac-Gr (solid curve) and zz-Gr (dash curve)
flakes  vs energy, and for 3 different aspect ratios. Note that
the number of quasi-periods is roughly inverse proportional to A
(i.e. proportional to the length), meaning that Fabry-Perot-like
resonances take place.} \label{GMR-A}
\end{figure}
%%%%%%%%%%%%%%%%%%%%%%%%%%%%%%%%%%%%%%%%%%%%%%%%%%%%%%%%%%%%%%%%%%%%%%%%%%%%%%

\section{Results}

We use Eqs.\ref{GMR} to determine the quantities of our main
interest. At first, for comparison, we have calculated GMR for zz-
and ac-Gr sheets \emph{vs.} energy which, in principle, is
proportional to the gate voltage. From the results in
Fig.\ref{GMR-A} one can see that GMR, at least in the ac-Gr case,
is quasi-periodic with the period-length roughly proportional to
the aspect ratio A, i.e inverse proportional to the length of the
graphene sheets L. Numerically, the periods are close to the
well-established value $p \cong 2$eV /L[nm],\cite{miao} or in the
present units $p \cong 0.8$ A/W[nm] (the energy unit is t=2.7 eV).
Incidentally, some features characteristic for Fabry-Perot
resonances in graphene have been already reported \cite{miao,
Ponomarenko}, where it has been also stressed that irregularities
and defects of graphene-edges may turn the conventional
Fabry-Perot picture into a spectacular quantum billiard-type one.
It is seen in Fig.~\ref{GMR-A} that the GMR factors for ac-Gr and
zz-Gr differ in 3 respects: (i) magnitude, (ii) the maxima of the
former are roughly equidistant in contrast to the latter, moreover
(iii) in the former the onset of the second sub-band is visible
(at $E \cong 0.08$, except for A=3.4, where $W>L$), whereas in the
latter it always is washed out. Qualitatively these results,
showing that GMR vs. energy (gate voltage) changes also in sign,
are consistent with the experiment of Ref. [\onlinecite{Cho}].

%%%%%%%%%%%%%%%%%%%%%%%%%%%%%%%%%%%%%%%%%%%%%%%%%%%%%%%%%%%%%%%%%%%%%%%%%%%%%
\begin{figure}[t]
\centering \epsfxsize=3.0in \epsfysize=3.2in
\rotatebox{0}{\epsfbox{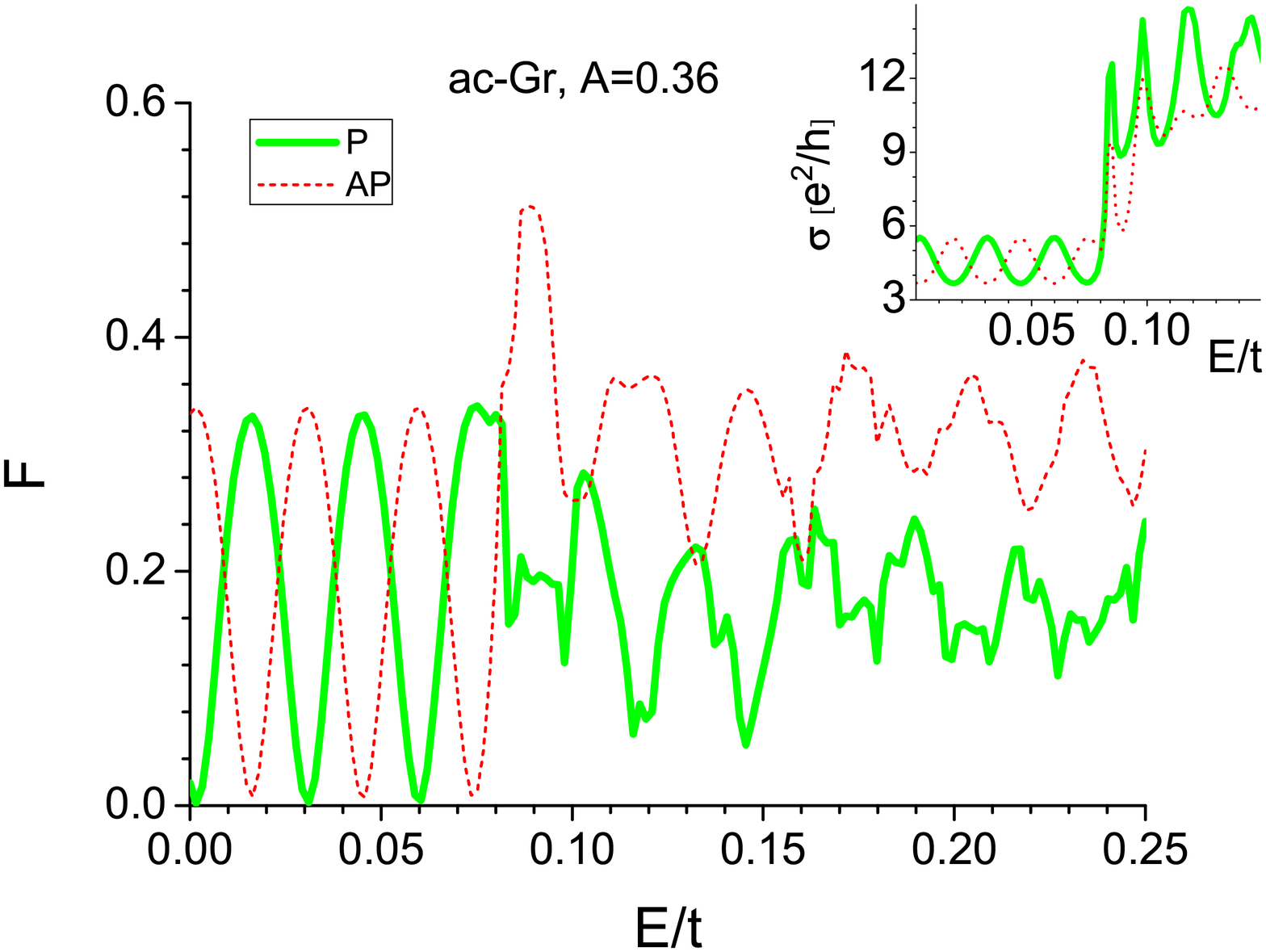}} \caption{ (Color online) Shot
noise Fano factor for the armchair graphene \emph{ca} 8nm in width
and with the aspect ratio A=0.36: for ferromagnetic electrodes
(parallel alignment - thin solid line, and antiparallel alignment
- dash line). Inset presents the respective conductivities.}
\label{FAC-54}
\end{figure}
%%%%%%%%%%%%%%%%%%%%%%%%%%%%%%%%%%%%%%%%%%%%%%%%%%%%%%%%%%%%%%%%%%%%%%%%%%%%%%

Another interesting point encountered here, is the universal
behavior issue due to evanescent modes at the Dirac points. It is
well-known that graphene sheets of large sizes, with $ W
\longrightarrow \infty $ and $ L \longrightarrow \infty $ (with
$W>L$), if homogenous, reveal universal values of $\textrm{\Large
$\sigma$}=4/\pi \, \, e^2 /h $ and
$F=1/3$.\cite{Tworzydlo,miao,Danneau}. The term "homogenous" in
this context means that the device at hand is an all-carbon
system, typically, with a central graphene sheet, and highly doped
graphene electrodes. This theoretical concept might be quite
realistic for widely applied experimental setups with
\emph{side-contacted} grahene sheets, provided evaporated metal
contacts are not invasive, i.e. they do not destroy the
underlaying honey-comb lattice but merely dope it slightly. It has
been also shown theoretically that in the case of
\emph{end-contacted} devices to 2-dimensional square-lattice
contacts, the situation is more complicated and respective values
of the conductivity and the Fano factor do depend on the on-site
potentials in electrodes and on details of bonds between
square-lattice electrodes and graphene.\cite{Dragomirova}
Importance of the interface conditions which determine, inter
alia, mixing of propagating modes, and whether or not the momentum
component along the interface direction is conserved, has been
noticed in Ref.[\onlinecite{Dragomirova, Blanter, Robinson, Lee,
Datta, Russo}]. It has been also demonstrated using one-parameter
scaling arguments that at the Dirac points, conductivity of an
infinitely large graphene with disorder is infinite (zero) if
there is not (there is) intervalley scattering. \cite{Bardarson}
Beyond the Dirac points the evanescent modes give way to the
propagating ones, and in the ballistic transport regime, perfect
noiseless transmission due to Fabry-Perot resonances may take
place.\cite{Cho,Cayssol}

Fig.\ref{FAC-54} clearly shows that this scenario also holds in
the ferromagnetic case for ac-Gr if the aspect ratio is not too
big (Gr ribbon is long enough). Indeed maxima in {\Large $\sigma$}
(Inset) correspond to minima of F (at F=0) in the main panel. In
the zz-Gr case with the same $A$ value the situation is similar
(not plotted), but then the oscillations in the P and AP
configurations are not phase-shifted.

\begin{figure}[t]
\centering \epsfxsize=3.0in \epsfysize=3.2in
\rotatebox{0}{\epsfbox{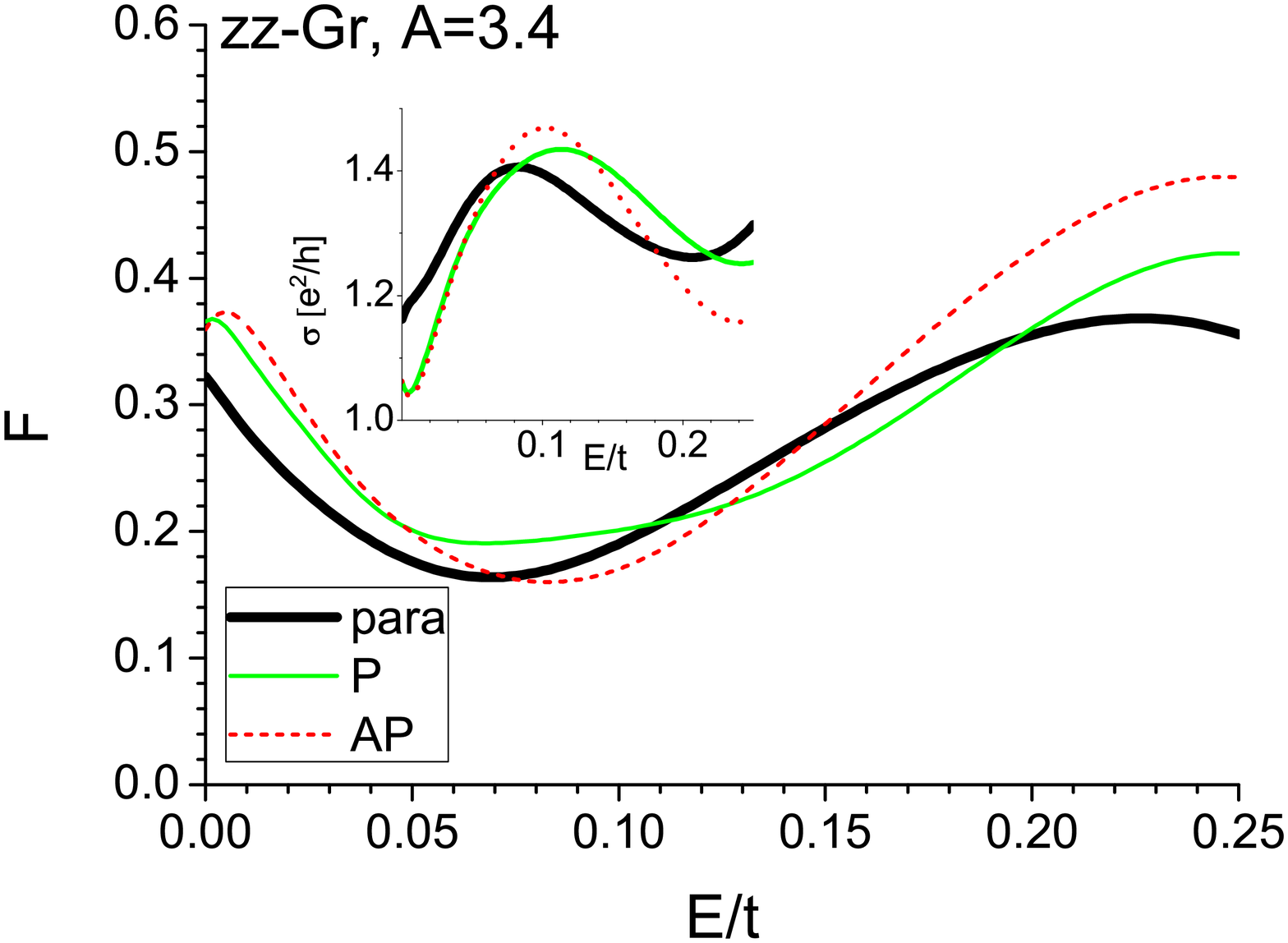}} \caption{ (Color online) Shot
noise Fano factor for the zigzag-edge graphene \emph{ca} 8nm in
width and with the aspect ratio A=3.4: for paramagnetic electrodes
(thick solid line), ferromagnetic electrodes (parallel alignment -
thin solid line, and antiparallel alignment - dash line). Inset
presents the respective conductivities.} \label{FZZ}
\end{figure}

\begin{figure}[t]
\centering \epsfxsize=3.0in \epsfysize=3.2in
\rotatebox{0}{\epsfbox{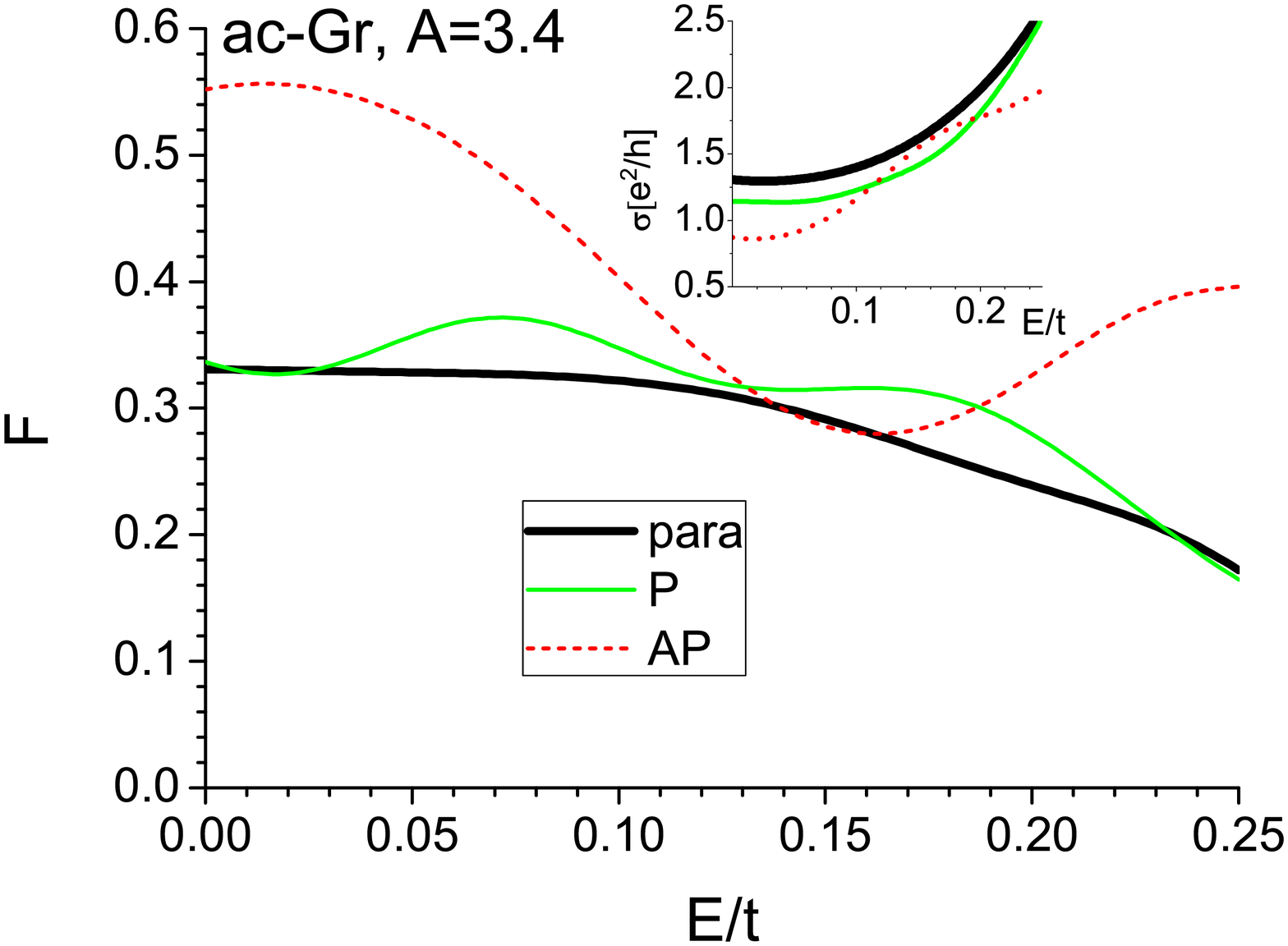}} \caption{ (Color online) As
\ref{FZZ} but for armchair-edge graphene. } \label{FAC}
\end{figure}

As regards Gr sheets with bigger $A$, Figs.\ref{FZZ} and \ref{FAC}
show some features of F and {\Large $ \sigma $} which happen to be
similar to the universal ones when the Gr sheets are
paramagnetically contacted (thick black lines). However, if the
contacts are ferromagnetic there is a tendency for F to increase
and for {\Large $\sigma$} -- to decrease. This suggests that in
general Gr-flakes studied here show non-universal behavior, which
is to be attributed to the finite sizes, and the inhomogeneity
resulting in contact-dependent charge transfer between interface
atoms (and accompanying electron-hole asymmetry). Incidentally,
the present approach was shown earlier to yield short-range charge
transfer, affecting mainly interface carbon
monolayers.\cite{krompiewski06,JPCM04} As a matter of fact, the
non-universal behavior of F and {\Large $\sigma$} in Gr-systems
has been already reported to be due to: charged impurity
scattering,\cite{Chen} strong disorder \cite{Lewenkopf}, invasive
contacts. \cite{Huard, Du}

Finally, it should be noted that the results presented in Figs.
2-5 hardly depend on the hopping parameters as long as the
interfaces are transparent enough. In the present theory this
condition is fulfilled provided the hopping parameter across the
interface ($t_c$) is not too different from the geometric mean of
hopping parameters for the metal electrodes ($t_M$) and graphene
($t$), i.e. from $t_c=\sqrt{t_M t}$. Otherwise, in case of a
drastic differentiation of the hopping parameters, and opaque
interfaces (very small $t_c$) the Coulomb blockade physics may
come into play,\cite{Weymann} out of reach of the present
approach.

%%%%%%%%%%%%%%%%%%%%%%%%%%%%%%%%%%%%%%%%%%%%%%%%%%%%%%%%%%%%%%%%%%%%%%%%%%%%%
%%%%%%%%%%%%%%%%%%%%%%%%%%%%%%%%%%%%%%%%%%%%%%%%%%%%%%%%%%%%%%%%%%%%%%%%%%%%%%

%%%%%%%%%%%%%%%%%%%%%%%%%%%%%%%%%%%%%%%%%%%%%%%%%%%%%%%%%%%%%%%%%%%%%%%%%%%%%%
%%%%%%%%%%%%%%%%%%%%%%%%%%%%%%%%%%%%%%%%%%%%%%%%%%%%%%%%%%%%%%%%%%%%%%%%%%%%%%

\section{Conclusion}

%%%%%%%%%%%%%%%%%%%%%%%%%%%%%%%%%%%%%%%%%%%%%%%%%%%%%%%%%%%%%%%%%%%%%%%%%%%%%%
%%%%%%%%%%%%%%%%%%%%%%%%%%%%%%%%%%%%%%%%%%%%%%%%%%%%%%%%%%%%%%%%%%%%%%%%%%%%%
%\begin{figure}[b]
%\centering \epsfxsize=3.2in \epsfysize=2.5in
%\rotatebox{0}{\epsfbox{Is_over_Ie.eps}} \caption{xxxx.}
%\label{NOS}
%\end{figure}
%%%%%%%%%%%%%%%%%%%%%%%%%%%%%%%%%%%%%%%%%%%%%%%%%%%%%%%%%%%%%%%%%%%%%%%%%%%%%%
%%%%%%%%%%%%%%%%%%%%%%%%%%%%%%%%%%%%%%%%%%%%%%%%%%%%%%%%%%%%%%%%%%%%%%%%%%%%%%

Summarizing, the aim of this study has been to estimate the effect
of the chirality as well as the aspect ratio on the GMR
coefficient of graphene flakes end-contacted to ferromagnetic
electrodes. In contrast to other theoretical approaches, the leads
are not supposed here to be two-dimensional (of honeycomb- or
square-lattice type), but more realistically they are modelled as
3-dimensional fcc-(111) semi-infinite slabs. It turns out that for
long and narrow systems (small aspect ratio), the GMR is a
quasi-periodic function of energy (gate voltage) with the period
roughly proportional to the aspect ratio, reflecting
Fabry-Perot-like resonances, typical of ballistic transport.
Notably, the GMR coefficient of the ac-Gr flakes may exceed 20\% -
40\%, depending on the aspect ratio value, whereas for zz-Gr
flakes the corresponding figures are distinctly smaller, but still
significant. The difference is due to the fact that in the case of
armchair-edge sheets all the interface carbon atoms belong to the
same sub-lattice (say, A-type), as opposed to the zigzag ones with
interface carbon atoms of both A- and B-type. This inevitably
facilitates intervalley mode mixing in the latter case. Another
noteworthy point, in the context of zero-energy (Dirac-point)
conductivity and Fano factor, is that for the big aspect ratio.
Such systems studied here show non-universal behavior when
ferromagnetic electrodes are applied. It is so even if the
corresponding values for the paramagnetic electrodes happen to be
close to the universal ones ($4/\pi \, \, e^2 /h $ and $1/3$ for
{\Large $\sigma$} and F, respectively). However, these systems are
inhomogeneous and rather small, so neither the limits $ W
\longrightarrow \infty $ and $ L \longrightarrow \infty $ (with
$W>L$), nor the requirement of a large number of propagating modes
can be fulfilled (see the relevant assumptions in
Ref[\onlinecite{Tworzydlo}]). It is noteworthy that the wide
end-contacted devices considered here resemble experimental setups
with invasive contacts, which do not show universal behavior,
either.

\section{Acknowledgments}
This work was supported by the EU FP6 grant CARDEQ under contract
No. IST-021285-2;
 and, as part of the European Science
Foundation EUROCORES Programme SPINTRA (contract No.
ERAS-CT-2003-980409), by the Ministry of Science and Higher
Education as a research project in 2006-2009.

\end{document}